\documentclass[conference]{IEEEtran}
\ifCLASSINFOpdf
\else
\fi
\usepackage{graphicx}
\usepackage{hyperref}
\usepackage{amssymb}
\usepackage{amsmath}
\graphicspath{{picture/}}

\begin{document}
%
\title{Design and optimization of DBSCAN Algorithm based on CUDA}



%
\author{\IEEEauthorblockN{Bingchen Wang,
Chenglong Zhang,
Lei Song,
Lianhe Zhao,
Yu Dou,
and Zihao Yu
}
\IEEEauthorblockA{ Institute of Computing Technology \\
	Chinese Academy of Sciences \\
	Beijing, China 100080
}}


\maketitle

\begin{abstract}
DBSCAN is a very classic algorithm for data clustering, which is widely used in many fields. However, with the data scale growing much more bigger than before, the traditional serial algorithm can not meet the performance requirement. Recently, parallel computing based on CUDA has developed very fast and has great advantage on big data. This paper summarizes the algorithms proposed before and improves the performance of the old DBSCAN algorithm by CUDA. The algorithm uses shared memory as much as possible compared with other algorithms and it has very good scalability. A data set is tested on the algorithm of new version. Finally, we analyze the results and give a conclusion that our algorithm is approximately 97 times faster than the serial version.
\end{abstract}


%
\IEEEpeerreviewmaketitle

\section{Introduction}

According to the data similarity, a group of data can be divided into some subsets. This progress is called clustering. Generally, there are four types of main cluster method, including partition methods, hierarchical methods, density-based methods and grid-based methods. DBSCAN \cite{1} has been widely used in many aspects as one of the density-based methods. Many methods has been proposed to improve the performance of DBSCAN by using MPI, MAP-REDUCE and other parallel computing methods \cite{2}\cite{3}.

In recent years, researches about General Purpose-Graphics Processing Unit(GPGPU) has been hot, especially the CUDA frame released by NVIDA has provided the support for C which gives much more convenience for the programming and improvement of GPU algorithms \cite{4}\cite{5}.GPU has many advantages in high performance computing such as high degree of parallelism and compute-intensive. Both of them are very suitable for data mining algorithms. Currently, some DBSCAN algorithms using parallel computing based on GPU has been proposed, including G-DBSCAN \cite{6},CUDA-DCLust \cite{7} and CUDASCAN \cite{8}. These algorithms are for more complicated with too many control flow branches. As a result, they have a low degree of parallelism and can not fully use the structure of GPU to improve the performance.

This paper divides the DBSCAN algorithm based on CUDA into three steps. 
\begin{enumerate}
	\item compute the distance between each pair of data-nodes and generate the distance matrix.
	\item get core-points by the distance matrix and generate primitive clusters
	\item merge the primitive clusters.
\end{enumerate}
Each step of the algorithm can be fully parallelized and easy to be improved according to the structure of GPU.  

The remaining of the paper is organized as five parts. The first part gives a brief introduction of the concepts about DBSCAN. The second part describes the DBSCAN algorithm based on GPU. The third part has given the details about how to improve the algorithm based on CUDA. The fourth part shows the experimental results and analysis. Finally, the conclusion of the research has been given.


\section{Background}

DBSCAN (Density-Based Spatial Clustering of Applications with Noise) is a data clustering algorithm based on density which can find out the high density connected area separated by those low density area. DBSCAN is not sensitive to the noises and is able to handle any kind of data in various shapes. \cite{9}

Here are some concepts about DBSCAN.

\begin{enumerate}
	\item $\epsilon$-Neighborhood: Given the radius $\epsilon$ of a point, all points within distance $\epsilon$ of it are called $\epsilon$-Neighborhoods.
	\item A point $p$ is a core point if at least $MinPts$ points are within distance $\epsilon$ of it, and those points are said to be directly reachable from $p$. No points are reachable from a non-core point.
	\item A point $q$ is reachable from $p$ if there is a path $p_1, \dots, p_n$ with $p_1 = p$ and $p_n = q$, where each $p_{i+1}$ is directly reachable from $p_i$ (so all the points on the path must be core points, with the possible exception of $q$).
\end{enumerate}

The definition of a cluster mainly contains two things:
\begin{enumerate}
	\item Given the cluster $C$, if $p\in C$ and $q$ is reachable from $p$ within distance $\epsilon$, then $q \in C$.
	\item $\forall p, q \in C$, $p$ is reachable from $q$. 
\end{enumerate}

Serial algorithm of DBSCAN contains three steps.
\begin{enumerate}
	\item Calculate the distance of each pair of data nodes and store them in a two-dimensional matrix¡£
	\item Build primitive clusters: find all core points and the neighborhood of each core point. The neighborhoods are primitive clusters.
	\item Merge primitive clusters: Find out all reachable core points by iteration and merge their primitive clusters.
\end{enumerate}

\section{Related Work}
DBSCAN algorithm based on GPU mainly includes G-DBSCAN, CUDA-DClust and CUDASCAN.

G-DBSCAN \cite{6} includes two steps: Firstly, generate the matrix of each pair of points according to eps. Secondly, begin to proceed BFS from any point and mark the points on the same tree as a cluster.

CUDA-DClust \cite{7} uses parallel computing of GPU blocks and proceed DFS at multiple points at the same time. Sub-clusters are called chains. By determining if there are overlapping core points between chains, the collision matrix is constructed and then chains are merged according to the collision matrix. 

CUDAScan algorithm \cite{8} divides a data set into domains stored in shared memory to compute. Each domain is calculated by a block to fully take the advantage of the speed of GPU visiting shared memory. Firstly, proceed DBSCAN in each domain to get local clusters separately and concurrently. Secondly, by determining if there are conflicts between local clusters to generate the collision matrix. Finally, merge all local clusters together.

The algorithms above mostly build local clusters with the help of tree structures. This method has a low degree of parallelism and difficult to deal with the cooperation of different blocks. So many global memory accesses consuming much time are in need. CUDAScan reduces visits for global memory through dividing the data. However, the size of blocks is limited and it is not suitable for high density data.

\section{Algorithm Design}

We use the serial algorithm running a test set with 23040 points and each point has a 3D coordinate. Then use the performance analysis tool \emph{gprof} to analyze the results.

\begin{table}[!htbp]
	\centering
	\caption{Execution times for each step of the serial DBSCAN algorithm with data set of size 23040}
	\label{1}
\begin{tabular}{|l|l|l|}
	\hline
	Step & Execution time/s & Ratio \\
	\hline
	Distance calculation & 1.63 & $66.27\%$ \\
	\hline
	primitive clusters construction & 0.80 & $32.62\%$ \\
	\hline
	Clusters merging & 0.03 & $1.22\%$ \\
	\hline
\end{tabular}
\end{table}

Table~\ref{1} shows that distance calculation consumes $66.27\%$ of total time which is the bottleneck of serial algorithm.

\subsection{Use CUDA to accelerate distance calculation}

The distance calculation of each pair of points only depends on the coordinate of both points which has nothing to do with other information. When all the coordinates of each points have been given, the distance of any two points can be calculated concurrently. So it is very easy to use CUDA to accelerate. Because the reason for distance calculation is to see if the distance between two points is shorter than $\epsilon$, distance calculation is faster if using $\epsilon^2$ to compare with instead of $\epsilon$.

\subsubsection{Baseline version}

Each CUDA thread is responsible for the distance calculation between one point and other $N$ points. Each thread fills in one line in the \texttt{distance} matrix. Threads go over the coordinate information of all points by loops which can calculate the distance between them and the goal point in each thread.

\subsubsection{Use memory coalescing}

In the baseline version, threads need to read the coordinate information from the point matrix in the global memory and write back the distance to \texttt{distance} matrix in the global memory. However both visits to global memory do not have the properties of memory coalescing. Transposing for both matrixes is done to improve the performance brought by memory coalescing. But we do not need to actually transpose the matrix. As for point matrix, we only need to change the definition of matrix(changed to \texttt{point[3][N]}) and the code about how to read the coordinate information by CPU. And \texttt{distance} matrix is symmetrical so that each thread writing one row in the matrix will get the same result. After changing, adjacent threads will visit the continuous positions in the point matrix at the same time. Meanwhile, distance can be written on continuous positions in the \texttt{distance} matrix to meet the need of memory coalescing. 

\subsubsection{Use shared memory}

In the last version, each thread will read coordinate information from global memory. In fact, every thread will use coordinate information of each points while calculating distance. Shared memory can be used for the sharing feature which is that each thread reads coordinate info of each points into shared memory so that each point only needs to be read once from the shared memory. It can be used by \texttt{TPB} threads in one thread block. Because the limitation of shared memory, we can not read info of all points into shared memory when \texttt{N} is too large. So we read them in different parts. Read coordinate information of \texttt{TPB} threads into shared memory each time and then calculate the distance between the goal point of each thread and \texttt{TPB} points. Finally, write results into \texttt{distance} matrix. Repeat the process until all points are visited. At the same time, a thread needs the coordinate information of thread goal point to calculate the distance, which is private for each thread. So we can store the information in the thread register before the distance calculation. As a result, the information about coordinate of thread goal points is only read once from the global memory which can be used in the next \texttt{N} times of distance calculation.

\subsubsection{Loop unrolling}

In the version of using shared memory, the inner loop only calculate one distance. We proceed the most inner loop unrolling, one iteration will calculate 32 distance information so that it will reduce control cost related to iterations.

\subsection{Use CUDA to construct primitive clusters}

We use \texttt{cluster} matrix to identify the relationship between the clusters and the points. Each column of the matrix means one cluster and each row means one point. For example, \texttt{cluster[i][j]=1} if and only if \texttt{point[j]} is in the \texttt{i}th cluster. Besides use the variable valid to identify if the cluster is valid. For example, \texttt{valid[i]=1} if and only if the cluster in ith column of \texttt{cluster} matrix is valid. The essence of the construction of primitive clusters is to construct \texttt{cluster} matrix. The specific process is as follows: for each point as in \texttt{point[i]}, scanning all the distance information with other points. If $\texttt{distance[i][j]}\le \epsilon^2$, then \texttt{cluster[i][j]} is set to $1$ which means the \texttt{point[j]} is located in the $\epsilon$-neighborhood of \texttt{point[i]}; During the scan, calculate the points which are in the $\epsilon$-neighborhood of \texttt{point[i]}.If the number is greater than $MinPts$, then \texttt{valid[i]} is set to $1$.

Constructing \texttt{cluster} matrix only depends on the distance information between all points. When the information is given, all columns of \texttt{cluster} matrix are independently constructed which can be built concurrently. So it is easy to use CUDA to accelerate. Each CUDA thread is responsible for writing one column of \texttt{cluster} matrix. The thread use the iteration to visit the corresponding column of \texttt{distance} matrix and to compare distance with $\epsilon^2$ which is recorded. Each CUDA thread is not for one element of the matrix but for a column. The reason for that is we need to count the number of points which are in the $\epsilon$-neighborhood to build the \texttt{valid} vector. The method will lead to extra operations while counting.

\subsubsection{Calculate distance and construct primitive clusters at the same time}

In fact, the \texttt{distance} matrix is only for constructing primitive clusters. The distance calculation and comparing the result with $\epsilon^2$ can be proceeded at the same time. Meanwhile construct the \texttt{cluster} matrix so that the distance do need to be stored in the global memory.

\subsubsection{Put the iteration code outsides}

In the last version, the most inner iteration of the thread is to calculate the square of Euclidean distance between thread goal point and other points. Assuming that the coordinate of the thread goal point is $(t_x,t_y,t_z)$ and the $n$th point is $(p[n]_x, p[n]_y, p[n]_z)$. Then the thread will need the following calculation.

$$dist(t, p[n]) = (t_x-p[n]_x)^2+(t_y-p[n]_y)^2+(t_z-p[n]_z)^2$$

Noticed that for a thread, the information of thread goal point do not change with the iteration. So we can move the calculation only related to thread goal point before the iteration. To do that, we need to change the way of distance calculation. 

\begin{align*}
 & dist(t, p[n]) \\
 = &t_x^2-2t_xp[n]_x+p[n]_x^2 + t_y^2-2t_yp[n]_y+p[n]_y^2 +\\
 &t_z^2-2t_zp[n]_z+p[n]_z^2 \\
 = &(t_x^2+t_y^2+t_z^2) -(2t_xp[n]_x+2t_yp[n]_y+2t_zp[n]_z)+ \\
 & (p[n]_x^2+p[n]_y^2+p[n]_z^2)
\end{align*}

Given $T, X, Y, Z, P[n]$ as follows:
\begin{align*}
	T &= t_x^2+t_y^2+t_z^2 \\
	X &= 2t_x \\
	Y &= 2t_y \\
	Z &= 2t_z \\
	P[n] &= p[n]_x^2+p[n]_y^2+p[n]_z^2
\end{align*}

We can see that $T, X, Y, Z$ are all available to calculate before the iteration and $P[n]$ is not relevant with the thread goal point. We can calculate after reading the information about \texttt{TPB} points and putting it into shared memory which is shared by all threads. So wee only need to calculate once about $p[n]$ and do not need to calculate at each thread.

After putting the iteration code outside, each thread only needs to calculate in the iteration as follows:

$$dist(t, p[n]) = T + P[n] - (Xp[n]_x+Yp[n]_y+Zp[n]_z)$$

\subsection{Use CUDA to merge clusters}

Cluster merging is a process which has strong dependence on order. Every merge changes the status of the cluster. When the ith cluster is merged with the jth cluster, points in the jth cluster will be added and the points in the ith cluster will be deleted. We use a \texttt{cluster} matrix to record a cluster which can helps to compute concurrently. When a point is added in one cluster, the corresponding element in the \texttt{cluster} matrix is set from 0 to 1. There are no conditions for elements changing from 1 to 0. This means that all clusters can be merged to one certain cluster at the same time without the need for synchronize or atomic operations. To delete a cluster, we only need to set corresponding component in the \texttt{valid} vector into 0.

Based on the analysis, we can use CUDA to accelerate the cluster merging easily. The kernel function will accept a target as one parameter to proceed target which means that all clusters are tried to merge with the target cluster. Each thread block is responsible for trying to merge one certain cluster with the target cluster and determining whether they can be merged. If the merge succeeds, set the corresponding component in the \texttt{valid} vector into 0.

\section{Performance Evaluation}

Configurations of GPU used for evalution are listed in Table~\ref{2}: 

\begin{table}[!htbp]
	\centering
	\caption{Configurations of GPU used for evalution}
	\label{2}
\begin{tabular}{|l|l|}
	\hline
	device name & Tesla K10.G2.8GB \\
	\hline
	totalGlobalMem & 3757637632 \\
	\hline
	sharedMemPerBlock & 49152 \\
	\hline
	regsPerBlock & 65536 \\
	\hline
	warpSize & 32 \\
	\hline
	memPitch & 2147483647 \\
	\hline
	maxThreadsPerBlock & 1024 \\
	\hline
	maxThreadsDim[3] & $1024 \times 1024 \times 64$ \\
	\hline
	maxGridSize[3] & $2147483647 \times 65535 \times 65535$ \\
	\hline
	totalConstMem & 65536 \\
	\hline
	device version & major 3, minor 0 \\
	\hline
	clockRate & 745000 \\
	\hline
	textureAlignment & 512 \\
	\hline
	deviceOverlap & 1 \\
	\hline
	multiProcessorCount & 8 \\
	\hline
\end{tabular}
\end{table}

\subsection{Evalution for distance calculation}

\begin{table}[!htbp]
	\centering
	\caption{Evalution for distance calculation with data size of 23040}
	\label{3}
	\begin{tabular}{|l|p{1cm}|p{1cm}|p{1.2cm}|}
	\hline
	kernel & Execution time/ms & Step Speedup & Cumulative Speedup \\
	\hline
	Baseline & 1482.24 & - & - \\
	\hline
	Use memory coalescing & 48.128 & 30.798 & 30.798 \\
	\hline
	Use shared memory & 23.867 & 2.0166 & 62.104 \\
	\hline
	Use loop unrolling & 5.3037 & 4.5000 & 279.47 \\
	\hline
\end{tabular}
\end{table}

From Table~\ref{3}, we can see that the performance improvement brought by memory coalescing is huge: Before we use memory coalescing, each thread of one warp has very far and different address while visiting the global memory so that we has to divide into 32 memory transaction to proceed. After we use memory coalescing, one wrap only needs one memory transaction. In theory, the performance has been improved 32 times. The test result shows that the performance is improved by 30.798 times which is very close to the theory number. Although loop unrolling does not reduce the actual calculating work, the performance improvement brought by it is very good. The complier can carry on static scheduling on the code and reduce the cost of loop control so that it will reduce the effect brought by pipeline blocking.

\subsection{Evaultion for building and merging primitive clusters}

\begin{table}[!htbp]
	\centering
	\caption{Evalution for building and merging primitive clusters with data size of 23040}
	\label{4}
	\begin{tabular}{|p{6cm}|p{1cm}|}
	\hline
	Kernel & Execution time/ms \\
	\hline
	Distance calculation (use unrolling loop) & 5.3037 \\
	\hline
	primitive clusters construction (use shared memory) & 44.848 \\
	\hline
	Distance calculation and primitive clusters construction at the same time & 25.336 \\
	\hline
	Put the iteration code outside & 12.502 \\
	\hline
	Merge clusters & 22.48 \\
	\hline
\end{tabular}
\end{table}

From Table~\ref{4}, we can see that if we separate the process of distance calculation and building primitive clusters, the time will be 50.151ms. But if we proceed the distance calculation and primitive clusters building at the same time, the time is only 25.336ms. The results of both methods are completely identical. It shows that half of the 50.151ms is for visiting global memory.

Besides, before loop code is put outside, it will need 8 floating point calculation to get one element of each \texttt{cluster} matrix. But after loop code is put outsides, the time is reduced to 6 floating point calculation. It is related to compiling results. Before loop code is put outsides, \emph{nvcc} compile about 442 instructions and 342 instructions after that. So we can get that the way of changing distance calculation has make \emph{nvcc} compile some more compact instructions.

Cluster merging is not particular ideal. The time consuming is equal to the serial algorithm. Most of the time is for data transmission.

\subsection{Overall Evalution}

\begin{table*}[t]
	\centering
	\caption{Overall evalution with different data size}
	\label{5}
\begin{tabular}{|p{1.2cm}|p{2.3cm}|p{2.3cm}|p{2.3cm}|p{2.2cm}|p{2.5cm}|p{2cm}|}
	\hline
	Data size & CPU time (distance calculation + primitive clusters construction)/ms & GPU time (distance calculation + primitive clusters construction)/ms & Speedup (distance calculation + primitive clusters construction) & CPU total time/ms & GPU total time (With data transmission)/ms & Overall Speedup \\
	\hline
	5061 & 130 & 0.8142 & 159.67 & 170 & 45.12 & 3.7677 \\
	\hline
	23040 & 1790 & 9.9077 & 180.67 & 1810 & 32.37 & 55.916 \\
	\hline
	60032 & 12180 & 69.323 & 175.70 & 12210 & 124.73 & 97.891 \\
	\hline
\end{tabular}
\end{table*}

From Table~\ref{5}, we can see that distance calculation and building primitive clusters are both bottleneck steps in the serial algorithm. CUDA acceleration has given a great performance improvement. However, CUDA acceleration brings difficulty of cluster merging. 

\section{Discussion}

At first, our plan was as following. First, calculate the distance of each pair of points. Second, get the primitive clusters and generate the primitive Boolean conflict matrix. Thirdly, calculate the transitive closure of Boolean conflict matrix and get global cluster numbers which can be merged by the same cluster markers. Fourth, re-write the result according to the global cluster number. The key step is to calculate the transitive closure which can be realized by Warshall algorithm. So every step has a high degree of parallelism. The plan transforms the original problem into a pure mathematic problem and the degree of parallelism of Warshall algorithm is very high. This seems a pretty good plan. However, the experimental results are not so good. Because even if Warshall algorithm has a high degree of parallelism, each calculation needs a launch of kernel which costs 3ms. And Warshall algorithm needs start Kernel for $N$ (the number of points) times.

The relationship of cluster and point is stored by matrix. At the worst case, the store method will need $O(N^2)$ space. And the scale of 60000 has reached the space limitation. Further study is still needed before we find out algorithm with good performance and scalability.

\section{Conclusion}

In fact, a few key codes in the algorithm cost most of the time in the whole program. So the key of improving the performance of the algorithm lies in how to deal with these hotspot. The limits of DBSCAN is how to calculate the distance of each pair of points. We separate the calculate process into many small and un-related processes by analyzing the relevance of the process. Meanwhile, we modify the serial algorithm, get primitive clusters and then merge clusters. The relevance of merge process has been reduced and the parallelization of the algorithm is also improved.

Besides the parallelization brought by multiple cores, algorithm improvement based on the structure of GPU can dramatically speed up single thread computing. Much time has been reduced on memory coalescing, bank conflict and branch divergency by analyzing data structure, diving the calculate process and controlling the memory accessing operations. Some skills used in the serial algorithm are also suitable for the improvement of CUDA. We modify the comparison between the distance and the threshold, as well as unrolling the loops. These skills brings very good results.

According to computing results, the execution speed of our algorithm is 97.891 times higher than the serial algorithm of DBSCAN on the condition of a large data scale.

\end{document}